\documentclass{PoS}

\usepackage{epsfig,rotating}
\usepackage{amsmath,amssymb}
\usepackage{amsfonts}
\usepackage{mathrsfs}
\usepackage{bbm}
\usepackage[normalem]{ulem}
\usepackage[T1]{fontenc}

\newtheorem{thm}{Theorem}[section]

\newtheorem{theorem}[thm]{Theorem}
\newtheorem{proposition}[thm]{Proposition}

\newtheorem{remark}[thm]{Remark}

\newcommand{\lsc}{\{\hspace{-0.1cm}[}			
\newcommand{\rsc}{]\hspace{-0.1cm}\}}

\newcommand{\CC}{\mathcal{C}}

\newcommand{\CH}{\mathcal{H}}

\newcommand{\CL}{\mathcal{L}}
\newcommand{\CM}{\mathcal{M}}

\newcommand{\CQ}{\mathcal{Q}}

\newcommand{\CV}{\mathcal{V}}

\newcommand{\CE}{\mathcal{E}}
\newcommand{\frg}{\mathfrak{g}}				

\newcommand{\frX}{\mathfrak{X}}

\newcommand{\FR}{\mathbbm{R}}     			
\newcommand{\NN}{\mathbbm{N}}     			
\newcommand{\RZ}{\mathbbm{Z}}     			

\newcommand{\dd}{\mathrm{d}}     			
\newcommand{\dpar}{\partial}     			

\newcommand{\de}{\mathrm{e}}     			

\newcommand{\eand}{{\qquad\mbox{and}\qquad}}     		

\newcommand{\der}[1]{\frac{\dpar}{\dpar #1}}   		
\newcommand{\ad}{\mathrm{ad}}     			

\newcommand{\sU}{\mathsf{U}}     			

\newcommand{\sG}{\mathsf{G}}
\newcommand{\sL}{\mathsf{L}}

\newcommand{\sO}{\mathsf{O}}

\newcommand{\sDiff}{\mathsf{Diff}}

\renewcommand{\remark}[1]{}     				
\def\tyng(#1){\hbox{\tiny$\yng(#1)$}}			
\def\tyoung(#1){\hbox{\tiny$\young(#1)$}}			

\newcommand{\beq}{\begin{eqnarray}}
\newcommand{\eeq}{\end{eqnarray}}

\title{Derived Brackets and Symmetries in Generalized Geometry and Double Field Theory}

\ShortTitle{Pre-N$Q$-manifolds and derived brackets in generalized geometry and double
field theory}

\author{\speaker{Andreas Deser} \\
  Istituto Nazionale di Fisica Nucleare - Sezione di Torino\\
  Via P.~Giuria 1\\
10125 Torino, Italy\\
        E-mail: \email{deser@to.infn.it}}

\author{Christian S{\"a}mann\\
       Department of Mathematics, Heriot-Watt University\\
Colin Maclaurin Building, Riccarton, Edinburgh EH14 4AS, U.K.\\
and\\
Maxwell Institute for Mathematical Sciences, Edinburgh,U.K. \\
and \\
Higgs Centre for Theoretical Physics, Edinburgh, U.K.\\
        E-mail: \email{c.saemann@hw.ac.uk}}

\abstract{Derived brackets as introduced and studied by Kosmann--Schwarzbach and Voronov are a powerful tool for describing and understanding infinitesimal symmetry actions relevant in physics. Roytenberg and Weinstein showed that this continues to hold for the categorified symmetries arising in Hitchin's generalized geometry. After reviewing some well-established examples, we prove that derived brackets also underlie the symmetries of Double Field Theory and heterotic Double Field Theory. This leads to a common framework for large classes of symmetries, which suggests that derived bracket constructions can function as a guiding principle in the description of infinitesimal actions of symmetries in physics. As a new result, we present sufficient conditions on a bracket to give rise to a Lie 2-algebra of symmetries via antisymmetrized derived brackets.}

\FullConference{Corfu Summer Institute 2017 "School and Workshops on Elementary
Particle Physics and Gravity"\\
                 2-28 September 2017\\
                 Corfu, Greece}

\begin{document}

\section{Introduction}

In symplectic geometry, the fundamental symmetry transformations are symplectomorphisms. These are called canonical transformations in classical physics and they leave physical observables invariant. At the infinitesimal level, one-parameter groups of symplectomorphisms are determined by the action of hamiltonian vector fields: Given a symplectic structure $\omega \in \Omega^2(M)$ on a manifold $M$ and hamilton function $f \in \mathcal{C}^{\infty}(M)$, the corresponding hamiltonian vector field $X_f$ is defined by $\iota_{X_f}\omega =\, \dd f$. On another observable $g \in \CC^{\infty}(M)$, the infinitesimal action of $X_f$ is determined by the Poisson bracket of $f$ and $g$. Using the Schouten--Nijenhuis bracket and the Poisson tensor $\pi \in \Gamma(\wedge^2 TM)$ corresponding to $\omega$, this can be written as
\begin{equation}
  \label{hamiltonaction}
  X_f(g) =\,\lbrace f,g\rbrace =\, \bigl[[-\pi,f],g\bigr]\;.
\end{equation}
This is the prime example of a \emph{derived bracket construction}~\cite{Kosmann-Schwarzbach:0312524, kosmann1996poisson}. One says that the Poisson bracket is a derived bracket by the Poisson tensor. 

Let us translate the last example into a more modern framework~\cite{Severa:2001aa}. Here, a Poisson manifold $(M,\pi)$ is equivalent to the degree shifted cotangent bundle\footnote{We will introduce parts of this language in the main text. For a detailed introduction we refer to standard textbooks on graded manifolds as well as reviews such as~\cite{Cattaneo:2010re}.} $T^*[1]M$ carrying a homological vector field $Q$ compatible with the canonical graded symplectic structure. Such a homological vector field has a Hamiltonian $\CQ$ which satisfies the relation $\lbrace \CQ, \CQ\rbrace_{T^*[1]M} =\, 0$. Observing that the exterior algebra $\Gamma(\wedge^\bullet TM)$ is isomorphic to polynomial functions on $T^*[1]M$ allows us to rewrite~\eqref{hamiltonaction} as
\begin{equation}
  \label{xfg}
  X_f(g) = \, \bigl\lbrace \lbrace \CQ,f\rbrace,g\bigr\rbrace\;,
\end{equation}
where we regard $f$ and $g$ as degree zero elements in $\CC^\infty(T^*[1]M)$. 

One of the advantages of this reformulation is its unifying power: It turns out that on any differentiable manifold, the action of one-parameter groups of diffeomorphisms on functions, given by the action of a vector field on a function, as well as the action of a vector field on another one can be written as a derived bracket on a suitably chosen graded manifold in the same form as~\eqref{xfg}. Even more: In case of Hitchin's generalized geometry, the fundamental symmetries are given by diffeomorphisms as well as closed $B$-field transformations ($B\in \Omega^2(M), \; dB=0$), i.e.~the group $\textrm{Diff}(M) \ltimes \Omega^2_{\textrm{cl}}(M)$~\cite{Hitchin:2010qz}. Their infinitesimal action on functions as well as on generalized vectors is determined by the Dorfman bracket, which is again a derived bracket as shown by Roytenberg and Weinstein~\cite{Roytenberg:1998vn, Roytenberg:1999aa, Roytenberg:0203110}. 

These examples are enough motivation for us to adopt the following guiding principle: \emph{Whenever we are seeking an infinitesimal action of a fundamental symmetry, we try to find the corresponding derived bracket description}. This principle is very successfully applied in the cases of Double Field Theory (DFT) and heterotic DFT~\cite{Deser:2014mxa, Deser:2017fko, Deser:2016qkw}, which further underlines its relevance. We thus expect that it will help uncover the action of infinitesimal symmetries in more complicated situations such as exceptional generalized geometries and exceptional field theories, especially for the so far problematic groups $E_7$ and $E_8$ or even beyond.

In the following, we shall explain the construction of the appropriate derived brackets for string gravity (gravity coupled to the Kalb--Ramond two-form $B$) as well as DFT and heterotic DFT. 

\section{$Q$-manifolds and pre-$Q$-manifolds}

A symmetry Lie group $\sG$ acting on a manifold of objects $X$ is intuitively described in terms of an {\em action Lie groupoid}. This is a very simple category with objects $X$ and morphisms parameterized by pairs $(g,x)\in \sG\ltimes X$, 
\begin{equation}
 x \xrightarrow{~(g,x)~}g\cdot x~.
\end{equation}
At an infinitesimal level, this translates into an action Lie algebroid, which is described as the trivial vector bundle $E=\frg\ltimes X\rightarrow X$ where $\frg$ is the Lie algebra of $\sG$. The Lie bracket on $\frg$ induces a Lie algebra structure on its sections and the action is encoded in a Lie algebra homomorphism $\rho:\Gamma(E)\rightarrow \Gamma(TX)$, known as the {\em anchor}. A vector bundle with this additional structure is called a {\em Lie algebroid}. 

Because supergravity and DFT involve higher form fields which are part of the connections of categorified circle bundles, we expect categorified Lie algebras and categorified Lie algebroids to play a crucial role. Fortunately, there is a very convenient language which allows for an accessible definition of both.

We start from a {\em smooth $\RZ$-graded manifold} $\CM$. Just as in the case of supermanifolds, we can picture $\CM$ as a $\RZ$-graded vector bundle $E$ over an ordinary manifold $\CM_0$ known as the {\em body} of $\CM$. If $(\xi^{a_i}_i)$ are the fiber coordinates\footnote{This is a rather simple minded picture of a locally ringed space, but it will suffice for our purposes; the $(\xi^{a_i}_i)$ are in fact the generators of the $\CC^\infty(\CM_0)$-sheaf of $\RZ$-graded rings.} on $E$ with $\xi^{a_i}_i$ of degree~$i$, then the algebra of functions $\CC^\infty(\CM)$ on $\CM$ is given by polynomials in the $\xi^{a_i}_i$ with coefficients in $\CC^\infty(\CM_0)$. A {\em homological vector field $Q$} on $\CM$ is a vector field of degree one satisfying $Q^2=0$. It is also a differential turning $\CC^\infty(\CM)$ into a differential graded algebra. A graded manifold with a homological vector field is a {\em $Q$-manifold}. If the vector bundle $E$ is trivial in negative degrees, we also speak of an {\em N$Q$-manifold}.

Let us consider two archetypical examples. First, let $\CM$ be the grade-shifted tangent bundle $T[1]M$ of some manifold $M$. Locally, we have coordinates $x^\mu$ of degree~0 on the base manifold $M$ and fiber coordinates $\xi^\mu$ of degree~1. A homological vector field is then $Q=\xi^\mu \der{x^\mu}$ and the differential graded algebra $(\CC^\infty(T[1]M),Q)$ is simply the de Rham complex $(\Omega^\bullet(M),\dd)$. The $Q$-manifold $\CM$ is also a Lie algebroid, namely the {\em tangent algebroid} of $M$.

Second, let $V$ be some vector space. The graded manifold $\CM=V[1]$ is a $\RZ$-graded vector bundle over a point which is non-trivial only in degree~1. We say that $\CM$ is {\em concentrated} in degree~1. Let $\xi^\alpha$ be the coordinates of degree~1 on $V[1]$. The most general homological vector field is of the form
\begin{equation}
 Q=-\tfrac12 f_{\alpha\beta}^\gamma\xi^\alpha\xi^\beta \der{\xi^\gamma}
\end{equation}
and the condition $Q^2=0$ translates into the Jacobi identity for the structure constants $f_{\alpha\beta}^\gamma$. Thus, a $Q$-manifold concentrated in degree~1 is a Lie algebra.

We can now readily generalize both examples and define the following. An {\em $L_\infty$-algebroid} is a $Q$-manifold and an {\em $L_\infty$-algebra} is a $Q$-manifold whose body is a point.

We shall consider some explicit examples of $L_\infty$-algebroids below. As an example of an $L_\infty$-algebra~\cite{Lada:1992wc}, consider the case of a Lie 2-algebra, which has an underlying N$Q$-manifold concentrated in degrees $1$ and $2$, $\sL=\sL_{1}\oplus \sL_2$. Analogously to the case of a Lie algebra regarded as an N$Q$-manifold, we indentify $\sL[-1]$ with the actual Lie 2-algebra $\frg=\frg_0\oplus \frg_1$. The homological vector field $Q$ now contains structure constants, which induce the following maps or {\em higher brackets}:
\begin{equation}
 \begin{gathered}
  \mu_1:\frg_1\rightarrow \frg_0~,\\
  \mu_2:\frg_0\wedge\frg_0\rightarrow \frg_0~,~~~\mu_2:\frg_0\times \frg_1\rightarrow \frg_1~,\\
  \mu_3:\frg_0\wedge \frg_0\wedge \frg_0\rightarrow \frg_1~.
 \end{gathered}
\end{equation}
The condition $Q^2=0$ translates into the {\em higher} or {\em homotopy Jacobi relations}, the first three of which read as 
\begin{equation}
\begin{aligned}
 \mu_1(\mu_2(w,v))&=\mu_2(w,\mu_1(v))~,~~~\mu_2(\mu_1(v_1),v_2)=\mu_2(v_1,\mu_1(v_2))~,\\
 \mu_1(\mu_3(w_1,w_2,w_3))&=\mu_2(w_1,\mu_2(w_2,w_3))+\mu_2(w_2,\mu_2(w_3,w_1))+\mu_2(w_3,\mu_2(w_1,w_2))~,\\
\end{aligned}
\end{equation}
for all $w,w_{1,2,3}\in \frg_0$ and $v,v_{1,2}\in \frg_1$. The first one states that $\mu_1$ is a differential, the second one is the compatibility relation of $\mu_1$ with $\mu_2$ and the last equation says that the violation of the Jacobi identity is controlled by an exact term given by $\mu_3$.

We will encounter some explicit examples of Lie 2-algebras later; more details on these can be found in the paper~\cite{Baez:2003aa}. Note that Lie $n$-algebras are then defined analogously.

To be able to define derived brackets, we shall need bracket structures on our $Q$-manifolds. More precisely, we shall be interested in Poisson brackets, mostly of non-zero degree\footnote{Poisson brackets of odd degree are sometimes also called {\em Gerstenhaber brackets}.}, which originate from symplectic forms. We will always require a symplectic form $\omega$ on a $Q$-manifold to be compatible with the homological vector field in the sense that $\CL_Q\omega=0$. In other words, $Q$ generates a symplectomorphism on the $Q$-manifold. If $\omega$ is of $\RZ$-degree~$n$, we speak of a {\em symplectic $Q$-manifold of degree~$n$}.

On a symplectic $Q$-manifold $(\CM,Q,\omega)$ of degree~$n$, each function $f\in \CC^\infty(\CM)$ comes with a Hamiltonian vector field $X_f$ of degree~$|f|-n$, defined implicitly via
\begin{equation}
 \iota_{X_f}\omega=\dd f~.
\end{equation}
The Poisson bracket between two functions $f,g\in \CC^\infty(\CM)$ is then given by
\begin{equation}
 \{f,g\}:=X_f g=\iota_{X_f}\dd g=\iota_{X_f}\iota_{X_g}\omega~.
\end{equation}
This Poisson bracket is of degree~$-n$ and graded antisymmetric, which amounts to the relations
\begin{equation}
 \begin{gathered}
 \{f,g\}=-(-1)^{(|f|+n)(|g|+n)}\{g,f\}~,\\
 \{f,gh\}=\{f,g\}h+(-1)^{(n-|f|)|g|}g\{f,h\}~,\\
 \{f,\{g,h\}\}=\{\{f,g\},h\}+(-1)^{(|f|+n)(|g|+n)}\{g,\{f,h\}\}
 \end{gathered}
\end{equation}
for all $f,g,h\in \CC^\infty(\CM)$. One can show~\cite{Roytenberg:0203110} that $Q$ itself is the Hamiltonian vector field of a function $\CQ$:
\begin{equation}
 Q f=X_{\CQ}f=\{\CQ,f\}~.
\end{equation}

Let us give a few examples of symplectic N$Q$-manifolds. A symplectic N$Q$-manifold of degree~$0$ is necessarily concentrated in degree~0, has trivial homological vector field and is thus an ordinary symplectic manifold. A symplectic N$Q$-manifold of degree~$1$ is necessarily of the form $T^*[1]M$ for some base manifold $M$ and the homological vector field encodes a Poisson tensor on $M$. 

An important class of examples is given by the {\em Vinogradov algebroids} $\CV_n$, cf.~\cite{MR1074539}, which are symplectic N$Q$-manifolds of degree~$n$ and given by the following data:
\begin{equation}
\CV_n(M):=T^*[n]T[1]M~,~~~\omega=\dd x^\mu\wedge\dd p_\mu +\dd \xi^\mu\wedge \dd \zeta_\mu\eand \CQ=\xi^\mu p_\mu~.
\end{equation}
Here, $x^\mu$ and $\xi^\mu$ are coordinates on the base and the fiber of $T[1]M$ of degrees~$0$ and~$1$, respectively, and $p_\mu$ and $\zeta_\mu$ are their duals in the cotangent fibers of degrees~$n$ and~$n-1$, respectively.

For the applications to Double Field Theory, it will turn out that we need to consider some relaxed form of symplectic $Q$-manifolds, in which the condition $Q^2=0$ is not imposed from the outset. We will refer to these as {\em symplectic pre-$Q$-manifolds}.

\section{Derived brackets and symplectic N$Q$-manifolds}

An important but rather unknown fact is now that a symplectic Lie $n$-algebroid comes with an {\em associated Lie $n$-algebra}, which is obtained from antisymmetrizing particular derived brackets. The details for the general case are found in~\cite{Roytenberg:1998vn,Fiorenza:0601312,Getzler:1010.5859,Ritter:2015ffa}; here, let us merely consider the case of a symplectic Lie $2$-algebroid $(\CM,Q,\omega)$ and its associated Lie 2-algebra $\frg$. If we decompose the functions on $\CM$ according to their $\RZ$-grading as $\CC^\infty(\CM)=\CC^\infty_0(\CM)\oplus\CC^\infty_1(\CM)\oplus\dots$, then the Lie 2-algebra consists of the vector spaces 
\begin{equation}
 \frg=\frg_0\oplus \frg_1=\CC^\infty(\CM)_1\oplus \CC^\infty_0(\CM)
\end{equation}
(note the exchange of grading), on which we define the higher brackets
\begin{equation}\label{eq:L_infty_brackets}
\begin{aligned}
\mu_1(\ell)&=Q\ell - \delta \ell~,\\
\mu_2(\ell_1,\ell_2)&=\tfrac12\big(\{\delta\ell_1,\ell_2\}\pm\{\delta\ell_2,\ell_1\}\big)~,\\
\mu_3(\ell_1,\ell_2,\ell_3)&=-\tfrac{1}{12}\big(\{\{\delta\ell_1,\ell_2\},\ell_3\}\pm \dots\big)~,
\end{aligned}
\end{equation}
where we abbreviated
\begin{equation}\label{def:delta}
 \delta(\ell)=\left\{\begin{array}{ll}
Q\ell & \ell\in \CC^\infty_1(\CM)~,\\
0 & \mbox{else}~.\\
\end{array}\right.\\
\end{equation}
Such Lie 2-algebras will prove to be the relevant to describe ordinary and categorified symmetries at an infinitesimal level.

We also note that this Lie 2-algebra $\frg$ has an action\footnote{By {\em an action of an $L_\infty$-algebra $\frg$ on some manifold $X$}, we mean a morphism of $L_\infty$-algebras from $\frg$ to the Lie algebra $\frX(X)$ of vector fields on $X$,~cf.~\cite{Mehta:2012ppa,Deser:2016qkw}.} on $\CC^\infty(\CM)$ via
\begin{equation}\label{eq:action_nu_2}
 \nu_2(\ell,f):=\{\delta \ell,f\}~.
\end{equation}
In particular, $\mu_2$ is simply the antisymmetrization of the restriction of the action $\nu_2$ to $\frg\subset \CC^\infty(\CM)$. This Lie 2-algebra action will be the relevant one for describing the infinitesimal actions on geometric objects.

\section{Examples: Lie- and Courant bracket as derived brackets}

Let us now spend some time illustrating the constructions introduced above by giving some examples. Actions of infinitesimal diffeomorphisms on manifolds or, more precisely, one-parameter groups of diffeomorphisms are determined by the Lie bracket of vector fields. The generalization of diffeomorphisms to the case of a principal $\sU(1)$-bundle is the semidirect product of diffeomorphisms and $\sU(1)$-gauge transformations. Finally, the action of the Lie algebra of $\Omega^2_{\textrm{cl}}(M) \ltimes \sDiff(M)$, whose elements are sections of the generalized tangent bundle, $TM \oplus T^*M$, is given by the Dorfman bracket of generalized geometry. All three cases can be recast in derived bracket form by choosing a suitable N$Q$-manifold.

\paragraph{The Lie bracket as derived bracket.} Let $M$ be a manifold. Then the Lie algebra of vector fields $X,Y$ is given by the associated Lie algebra of $\CM = \CV_1(M)$. Its action on functions and general polyvector fields on $M$ is given by $\nu_2$ from~\eqref{eq:action_nu_2}, as we shall show now. In a local chart of $\cal M$, we have degree zero coordinates $x^\mu, \zeta_\mu$ and degree one coordinates $\xi^\mu, p_\mu$. The homological vector field $\CQ = \xi^\mu p_\mu$ trivially squares to zero and for vector fields $X = X^\mu \zeta_\mu$ and $Y=Y^\nu \zeta_\nu$ we have
\begin{equation}
 \mu_2(X,Y)=\tfrac12\big(\{QX,Y\}-\{QY,X\}\big)=X^\mu \dpar_\mu Y^\nu\zeta_\nu-Y^\mu\dpar_\mu X^\nu\zeta_\nu=[X,Y]~.
\end{equation}
Note that functions $f$ on $\CM$ depending exclusively on $x$ and $\zeta$ can be identified with polyvector fields. The action of the Lie derivative on these functions is then given by $\nu_2$ from~\eqref{eq:action_nu_2}:
\begin{equation}
 \CL_X f=\{QX,f\}~.
\end{equation}
This action can be further extended to the Lie derivative of more general tensors such as the metric in case of Riemannian geometry~\cite{Deser:2016qkw}.

\paragraph{Principal $U(1)$-bundles.} The above picture can be extended to the symmetries of principal $\sU(1)$-bundles over $M$. Here, the symmetries of observables are locally the semidirect product of $\sU(1)$-gauge transformations and diffeomorphisms on $M$. The corresponding Lie algebra has elements of the form $f + X$ where $X$ is a vector field on $M$ and $f$ is a $\frak{u}(1)$-valued function on $M$. The Lie bracket is again given by 
\begin{equation}\label{principalbundle}
 \mu_2(f+X,g+Y)=\tfrac12\big(\{QX,Y\}-\{QY,X\}\big)=[X,Y] + \CL_X g - \CL_Y f\;,
\end{equation}
and a Lie derivative is defined by
\begin{equation}
 \CL_X f=\{QX,f\}
\end{equation}
(as well as extensions, cf.~\cite{Deser:2016qkw}). Again, both arise from derived brackets on $\CV_1(M)$, where the generalized vectors are the degree zero objects coming from functions on $M$ and functions on $\CV_1(M)$ linear in $\zeta^\mu$.

\paragraph{The Courant bracket.} Finally we recall the observation of Roytenberg and Weinstein~\cite{Roytenberg:1998vn,Roytenberg:1999aa} that the bracket on a Courant algebroid is a derived bracket. Let $M$ be the base manifold and consider the Vinogradov algebroid $\CM=\CV_2(M)$. Recall our choice of local coordinates $(x^\mu,\xi^\mu,\zeta_\mu,p_\mu)$ of degrees $(0,1,1,2)$. Extended vector fields, i.e.~objects of degree one, correspond to generalized vectors in the sense of Hitchin. Taking $X + \alpha = X^\mu \zeta_\mu + \alpha_\mu \xi^\mu $ and $Y + \beta = Y^\mu \zeta_\mu + \beta_\mu \xi^\mu$ and the homological vector field $\CQ = \xi^\mu p_\mu $, we observe that~\eqref{eq:L_infty_brackets} translates to
\begin{equation}
  \begin{aligned}
    \mu_1(f) &=\; \dd f\;,\\
    \mu_2(X+\alpha,f) &=\;\tfrac{1}{2}\,\iota_X\,\dd f\;,\\
    \mu_2(X+\alpha,Y+\beta) &=\; [X,Y] + \CL_X \beta - \CL_Y\alpha - \tfrac{1}{2}\,\dd(\iota_X\beta - \iota_Y\alpha)\;.
  \end{aligned}
  \end{equation}
Thus, $\mu_2$ gives the Courant bracket of generalized geometry. Moreover, the action~\eqref{eq:action_nu_2} reads as
\begin{equation}
 \nu_2(X+\alpha,Y+\beta) =[X,Y]+\CL_X\beta-\iota_Y\dd\alpha~,
\end{equation}
the Dorfman bracket of generalized geometry. Again, we see that the action of infinitesimal generalized diffeomorphisms is of derived bracket form.

To sum up the three examples we have strong motivation to take the derived bracket construction as \emph{guiding principle} to uncover the underlying algebraic structure of infinitesimal actions of generalizations of diffeomorphisms motivated by physical theories. In the following chapter we will see that this works for the C- and D-brackets of DFT, where generalized diffeomorphisms include ordinary diffeomorphisms as well as transformations in the direction of closed string winding coordinates. 

\section{The C- and D-brackets as derived brackets}

Closed string (field) theory\footnote{$L_\infty$-structures arising in field theories inspired by string field theory were explored in \cite{Hohm:2017pnh, Hohm:2017cey} and for W-algebras in~\cite{Blumenhagen:2017ogh, Blumenhagen:2017ulg}.} on backgrounds of the form $\FR^{n-1,1}\times T^d$, where $T^d$ is the d-dimensional torus has observables depending on the usual momentum degrees of freedom on the flat factor plus a discrete momentum and winding degree of freedom for every torus coordinate. Restricting the discussion to the tours part, $M=T^d$, a Fourier transform gives a configuration space which contains the torus coordinates plus a set of winding coordinates, i.e.~locally it is of the form $M \times \tilde M$. DFT is a field theory on this doubled configuration space \cite{Geissbuhler:2013uka,Hohm:2013bwa} and the field variables are the ones of the massless sector of closed string theory, i.e.~a dilaton $d$, a metric $g$ and a $B$-field. The latter two are combined into the generalized metric $\CH$. Choosing a splitting of $M\times \tilde M$ into two factors, it has the form of the metric on the generalized tangent bundle $TM \oplus T^*M$:
\begin{equation}
  \CH_{MN}=\left(\begin{array}{cc} g_{\mu\nu}-B_{\mu\kappa}g^{\kappa\lambda}B_{\lambda\nu}  &  B_{\mu\kappa}g^{\kappa\nu}\\ -g^{\mu\kappa}B_{\kappa\nu}  &g^{\mu\nu}  \end{array}\right)\;.
\end{equation}
For the dilaton $d(x,\tilde x)$ and metric $\CH(x,\tilde x)$, an action principle was found~\cite{Hohm:2010jy} which contains the action of ordinary bosonic low energy supergravity in the limit where fields do not depend on the dual coordinates:
\begin{equation}
\begin{aligned}
 S=\int_M \de^{-2d}\,\dd x^1\wedge \dots \wedge \dd x^D \Big(& \tfrac18 \CH^{MN}\dpar_M\CH^{KL}\dpar_N \CH_{KL}-\tfrac12 \CH^{MN}\dpar_M\CH^{KL}\dpar_L\CH_{KN}\\
 &~~-2 \dpar_Md \dpar_N \CH^{MN}+ 4\CH^{MN}\dpar_M d\dpar_N d\Big)~.
\end{aligned}
\end{equation}
This action is invariant under the structure group\footnote{This is defined by matrices $A\in \textrm{Mat}(2d)$, satisfying $A^t \eta A = \eta$, where $\eta =\bigl( \begin{smallmatrix} 0&\textrm{id}\\ \textrm{id}&0 \end{smallmatrix} \bigr)$.} $\sO(d,d)$  of generalized geometry on $M$ and has a local gauge symmetry whose algebra of infinitesimal transformations we are interested in. It turns out~\cite{Hohm:2010pp} that these infinitesimal transformations are parameterized by generalized vectors as in the third example of the previous chapter, but now depending on all of the coordinates of $M\times \tilde M$. Using the fundamental representation of $\sO(d,d)$, we can introduce the coordinates $X^M:= (X^\mu,\alpha_\mu)$ as well as the partial derivatives $\partial_M := (\partial_\mu,\tilde \partial^\mu)$. In this notation, the C-bracket reads
\begin{equation}\label{eq:Courant_bracket}
 \Bigl([X,Y]_C\Bigr)^M :=\,X^K\partial_K Y^M - Y^K\partial_K X^M -\tfrac{1}{2}\Bigl(X^K\partial^M Y_K - Y^K\partial^M X_K\Bigr)\;.
\end{equation}

We now wish to identify this C-bracket as a derived bracket, analogously to the Courant bracket on the Courant algebroid. As a local model for $M\times \tilde M$, we take $T^*M$. This choice is common in the literature, especially in the description of non-geometric backgrounds~\cite{Mylonas:2012pg, Mylonas:2013jha, Aschieri:2015roa}. The Courant algebroid example gives the idea: We start with $\CV_2(T^*M)$ with local coordinates $(x^M,p_M,\xi^M,\zeta_M)$ = $(x^\mu,\tilde x_\mu,...,\zeta_\mu,\tilde \zeta^\mu)$. From the first chapter, we know that this has a canonical symplectic structure and homological vector field given by
\begin{equation}
  \label{dftpoisson}
 \omega=\dd x^M\wedge \dd p_M+\dd \xi^M\wedge \dd \zeta_M~,~~~\CQ=\sqrt{2}\xi^M p_M~,
\end{equation}
the scale factor in $\CQ$ is introduced for later convenience. Looking at degree one, we notice that we have too many generalized vectors. Using the metric $\eta$, we define the following combinations
\begin{equation}
 \theta^M=\frac{1}{\sqrt{2}}(\xi^M+\eta^{MN}\zeta_N)\eand \beta^M=\frac{1}{\sqrt{2}}(\xi^M-\eta^{MN}\zeta_N)~.
\end{equation}
and drop all the dependence on $\beta$. This reduces the symplectic structure and $\CQ$ of~\eqref{dftpoisson} to
\begin{equation}
 \omega=\dd x^M\wedge \dd p_M +\tfrac12\eta_{MN}\dd \theta^M\wedge \dd \theta^N~,~~~\CQ=\theta^M p_M~.
\end{equation}
We observe that $\lbrace \CQ,\CQ\rbrace \neq 0$, but we still have $\CL_Q \omega = 0$ for the hamiltonian vector field $Q$ corresponding to $\CQ$. Thus we arrive at a symplectic pre-$NQ$-manifold $\CE_2(M):=\,(T^*[2] \oplus T[1])(T^*M)$, obtained as a reduction of $\CV_2(T^*M)$. 

In order to recover a finite term $L_\infty$-algebra structure as infinitesimal symmetries by formulas~\eqref{eq:L_infty_brackets}, we have to restrict the algebra of functions in degree zero and one to a subset $\sL(\CE_2(M)) \subset \CC^\infty (\CE_2(M))$:
\begin{theorem}
  The elements of the subset $\sL(\CE_2(M)) \subset \CC^\infty(\CE_2(M))$ of degree zero and one form a Lie 2-algebra with higher brackets~\eqref{eq:L_infty_brackets}, if the Poisson brackets close on $\sL(\CE_2(M))$ and the following conditions are satisfied for all $f,g\in\sL_0(\CE_2(M))$ and $X = X_M \theta^M,Y = Y_M\theta^M, Z = Z_M \theta^M\in\sL_1(\CE_2(M))$:
  \begin{equation}\label{eq:restrictions_LM}
  \begin{gathered}
    \{Q^2 f,g\}+\{Q^2 g,f\}=\;2\partial_M f\;\partial^M g=\;0~,\\
    \{Q^2 X,f\}+\{Q^2 f,X\}=\;2\partial_M X \;\partial^M f=\;0~,\\
    \{\{Q^2X,Y\},Z\}_{[X,Y,Z]}=2\theta^L\big((\dpar^MX_L)(\dpar_MY^K)Z_K\big)_{[X,Y,Z]}=0~,
  \end{gathered}
  \end{equation}
  where we use the metric $\eta_{MN}$ to raise and lower indices, i.e.~$\partial^M :=\,\eta^{MN}\partial_N$.
\end{theorem}
We observe that the first two conditions of the theorem give the strong section condition of DFT on functions and on functions and vector fields, respectively. This is a purely algebraic motivation for this constraint, independent of its physical origin as a weakening of the level matching condition of closed string theory. The third condition of the previous theorem is a weakening of the strong section condition, if only vector fields are considered: Indeed, the condition is fulfilled if the section condition holds due to the contraction of partial derivatives. This weakening plays a role in setting up Riemannian geometry for DFT, and we refer the reader to the original paper~\cite{Deser:2016qkw} for details on this point.

The full Lie 2-algebra structure on a subset $\sL(\CE_2(M))=\sL_0^\infty(\CE_2(M)) \oplus \sL_1^\infty(\CE_2(M))$ of\linebreak $\CC^\infty(\CE_2(M))$ for which conditions \eqref{eq:restrictions_LM} hold is then given by the higher brackets 
  \begin{equation}
    \begin{aligned}
      \mu_1(f)&=\; \lbrace \CQ,f\rbrace =:\;Qf=\theta^M\dpar_M f~,\\
       \mu_2(X,f)&=-\mu_2(f,X)=\tfrac12\{QX,f\}=\tfrac12 X^M\dpar_M f~,\\
  \mu_2(X,Y)&=-\mu_2(Y,X)=\tfrac12\big(\{QX,Y\}-\{QY,X\}\big)\\
  &=X^M\dpar_M Y-Y^M\dpar_M X+\tfrac12\theta^M(Y^K\dpar_MX_K-X^K\dpar_MY_K)~,\\
  \mu_3(X,Y,Z)&=\tfrac13\big(\{\mu_2(X,Y),Z\}+\{\mu_2(Y,Z),X\}+\{\mu_2(Z,X),Y\}\big)\\
  &=X^MZ^N\dpar_MY_N-Y^MZ^N\dpar_MX_N+Y^MX^N\dpar_MZ_N\\
  &\hspace{1cm}-Z^MX^N\dpar_MY_N+Z^MY^N\dpar_MX_N-X^MY^N\dpar_MZ_N~.
    \end{aligned}
    \end{equation}
Here, $f$ is a function on the base, interpreted as degree zero element in $\CC^{\infty}(\CE_2(M))$ and $X=X_M\theta^M, Y=Y_M\theta^M, Z=Z_M\theta^M$ are generalized (extended) vector fields, i.e.~degree one objects in $\CC^\infty(\CE_2(M))$.

This picture clarifies the algebraic structure of infinitesimal symmetry transformations in DFT. Furthermore, we observe that the C-bracket is given by the operation $\mu_2$ when acting on degree one objects. 

The D-bracket between generalized vectors is again given by the action~\eqref{eq:action_nu_2}. This action is readily generalized to arbitrary tensors in DFT as defined in~\cite{Deser:2016qkw}. These include in particular the generalized metric $\CH_{MN}$.

\section{Generalizations}

Let us finish our discussion of derived brackets with a new result. So far, the derived brackets we encountered were obtained from Poisson brackets on N$Q$- and pre-N$Q$-manifolds. In the following, we develop a sufficient criterion for a general bracket to lead to a Lie 2-algebra of symmetries via the derived bracket construction. An explicit example of such a general bracket is given by the $\alpha'$-deformed Poisson bracket underlying heterotic generalized geometry and Double Field Theory~\cite{Deser:2014wva,Deser:2017fko}.

We start from an $\NN$-graded manifold $\CM$ concentrated in degrees 0, 1 and 2, together with a function $\CQ$ of degree~3 and a bilinear bracket
\begin{equation}
 \lsc -,- \rsc~:~\CC^\infty(\CM)\times \CC^\infty(\CM)\rightarrow \CC^\infty(\CM)
\end{equation}
of degree~$-2$. Then we have the following theorem.

\begin{theorem}
  \label{deformedp}
 The graded vector space 
\begin{equation}
 \frg=\frg_0\oplus \frg_1=\CC^\infty_1(\CM)\oplus \CC^\infty_0(\CM)
\end{equation}
 together with the antisymmetrized derived brackets 
\begin{equation}\label{eq:L_infty_brackets2}
\begin{aligned}
\mu_1(\ell)&=\lsc \CQ,\ell\rsc - \delta(\ell)~,\\
\mu_2(\ell_1,\ell_2)&=\tfrac12\big(\lsc\delta\ell_1,\ell_2\rsc\pm\lsc\delta\ell_2,\ell_1\rsc\big)~,\\
\mu_3(\ell_1,\ell_2,\ell_3)&=-\tfrac{1}{12}\big(\lsc\lsc\delta\ell_1,\ell_2\rsc,\ell_3\rsc\pm \dots\big)
\end{aligned}
\end{equation}
with
\begin{equation}\label{def:delta}
 \delta(\ell)=\left\{\begin{array}{ll}
\lsc \CQ,\ell\rsc & \ell\in \CC^\infty_0(\CM)~,\\
0 & \mbox{else}~,\\
\end{array}\right.\\
\end{equation}
forms a Lie 2-algebra, if the following conditions are satisfied:
\begin{equation}\label{eq:conditions_Lie_2}
\begin{gathered}
 \lsc f,g\rsc=-(-1)^{(2+|f|)(2+|g|)}\lsc g,f\rsc~,\\
 \lsc \CQ,\lsc \CQ,f,\rsc\rsc=0~,\\
 \lsc \CQ,\lsc f,g \rsc\rsc=\lsc\lsc \CQ,f\rsc,g\rsc+(-1)^{2+|f|}\lsc f,\lsc \CQ,g\rsc\rsc~,\\
 \lsc X,\lsc \lsc \CQ,Y\rsc,\lsc \CQ,Z\rsc\rsc\rsc=\lsc \lsc X,\lsc \CQ,Y\rsc\rsc,\lsc \CQ,Z\rsc\rsc+\lsc \lsc \CQ,Y\rsc,\lsc X,\lsc \CQ,Z\rsc\rsc\rsc~,\\
 \lsc\lsc \lsc \CQ,X\rsc,\lsc \CQ,Y\rsc\rsc,Z\rsc=\lsc \lsc \CQ,X\rsc,\lsc\lsc \CQ,Y\rsc,Z\rsc\rsc+\lsc\lsc \lsc \CQ,X\rsc,Z\rsc,\lsc \CQ,Y\rsc\rsc~,\\
 \lsc X,\lsc \lsc \CQ,Y\rsc,Z\rsc\rsc=\lsc \lsc X,\lsc \CQ,Y\rsc\rsc,Z\rsc+\lsc \lsc \CQ,Y\rsc,\lsc X,Z\rsc\rsc
\end{gathered}
\end{equation}
for all $f,g\in \CC^\infty(\CM)$, $X,Y,Z\in \CC^\infty_1(\CM)=\frg_0$. Here, $|f|$ denotes the degree of $f$ as an element of $\CC^\infty(\CM)$.
\end{theorem}

The proof for this theorem is a cumbersome but straightforward insertion of the antisymmetrized derived brackets~\eqref{eq:L_infty_brackets2} into the higher homotopy Jacobi relations of a Lie 2-algebra. Using the conditions~\eqref{eq:conditions_Lie_2}, the latter become identities.

To show that this theorem is non-trivial, we should give at least one relevant example. Consider again the Vinogradov algebroid $\CV_2(M)=T^*[2]T[1]M$ with coordinates $(x^\mu,\xi^\mu,\zeta_\mu,p_\mu)$ of degrees~$0,1,1,2$, respectively and symplectic form and Hamiltonian given by
\begin{equation}
 \omega=\dd x^\mu\wedge \dd p_\mu+\dd \xi^\mu\wedge \dd \zeta_\mu~,~~~\CQ=\xi^\mu p_\mu~.
\end{equation}
Then we can make the following statement.

\begin{proposition}
The deformed Poisson bracket 
\begin{equation}\label{eq:def_Poisson_GG}
 \lsc f,g \rsc=\{f,g\}+\alpha'\left(f \overleftarrow{\der{x^\mu}}\overleftarrow{\der{\zeta_\nu}}\overrightarrow{\der{x^\nu}}\overrightarrow{\der{\zeta_\mu}}g\right)~.
\end{equation}
satisfies the conditions~\eqref{eq:conditions_Lie_2} and therefore gives rise to a Lie 2-algebra structure on $\frg=\CC^\infty_1(\CM)\oplus \CC^\infty_0(\CM)$ via antisymmetrized derived brackets.
\end{proposition}
The proof follows by straightforward computation: First, we readily observe that 
\begin{equation}
\lsc f,g\rsc=-(-1)^{(2+|f|)(2+|g|)}\lsc g,f\rsc\eand \lsc \CQ,f\rsc=\{\CQ,f\}~.
\end{equation}
The latter straightforwardly implies
\begin{equation}
 \lsc \CQ,\lsc f,g \rsc\rsc=\lsc\lsc \CQ,f\rsc,g\rsc+(-1)^{2+|f|}\lsc f,\lsc \CQ,g\rsc\rsc~.
\end{equation}
The checks of the remaining special cases of the Jacobi identity for $\lsc-,-\rsc$ are again tedious but straightforward.

We note that the deformed Poisson bracket~\eqref{eq:def_Poisson_GG} governs heterotic generalized geometry in the reduced picture, where the generalized tangent bundle is restricted from $TM\oplus \ad_{\frg}\oplus T^*M$ to $TM\oplus T^*M$. In particular, we have the pairing, the generalized Lie derivative or Dorfman bracket and the Courant bracket:
\begin{equation}
 \begin{aligned}
  \langle X,Y\rangle_{\alpha'}&=\lsc X,Y\rsc~,\\
  \hat \CL^{\alpha'}_X Y&=\lsc \lsc \CQ ,X\rsc ,Y\rsc~,\\
  [X,Y]_C^{\alpha'}&=\tfrac12\big(\lsc\lsc\CQ,X\rsc,Y\rsc-\lsc\lsc\CQ,Y\rsc,X\rsc\big)~.
 \end{aligned}
\end{equation}
For further details on this particular description, see~\cite{Deser:2017fko}.

It is now an obvious question whether the above generalizes to heterotic Double Field Theory. And indeed, the guiding principle of derived brackets works as expected and we find a deformation of the Poisson bracket on a pre-N$Q$-manifold, which describes the pairing, the D-bracket and the whole Lie 2-algebra of symmetries of heterotic Double Field Theory, cf.~\cite{Deser:2014wva,Deser:2017fko}.

\section{Conclusion}

We outlined the significance of derived bracket constructions in finding the action of infini\-tesimal symmetries on various physically motivated configuration spaces. The infinitesimal action is given by a derived bracket in all instances and the corresponding Lie-, Courant, or binary $L_\infty$-algebra bracket is its antisymmetrization. The latter is part of the structure maps of a Lie $n$-algebra, which we claim to be a unifying description of infinitesimal symmetries. 

These structures continue to exist even if we consider derived brackets with respect to \emph{deformed} Poisson structures as is the case in $\alpha'$-deformed generalized geometry (and DFT) for the heterotic string. Theorem \ref{deformedp} lists the necessary conditions for such deformations to give Lie 2-algebras.  

Clearly, our construction should next be applied to exceptional generalized geometries obtained from toroidal compactification of eleven dimensional supergravity \cite{Hull:2007zu, Berman:2011pe, Berman:2011cg, Coimbra:2011ky, Coimbra:2012af, Berman:2013eva} . In case of compactification on a four dimensional torus, the corresponding higher brackets of the underlying Lie 3-algebra are described by the antisymmetrized derived brackets on the Vinogradov algebroid $\CV_3$. For compactification on higher dimensional tori, the construction is more involved. However, it was again shown in \cite{Baraglia:2011dg} that the higher brackets are of a derived bracket type. The general formulation in terms of Poisson structures on certain graded manifolds remains an open problem (as seems to be the structure of the generalized tangent bundle). 

The extended field theories, which are the analogues of DFT for 11d supergravity are far less understood. Here, the derived bracket construction might be a first tool to actually identify the correct tangent bundle structure as well as the action of infinitesimal symmetries. Going beyond the group $E_7$ is an even more ambitious goal.

We want to conclude with a remark about quantization. Whenever an action of an infinitesimal symmetry can be written in derived bracket form via Poisson brackets on a suitable (graded) Poisson manifold, it is tempting to interpret it as semi-classical limit of some quantized version. However, the phase space which is quantized in our case is not the canonical phase space consisting of position and momenta obtained from the underlying field theory. Instead it is the graded Poisson algebra corresponding to the exterior algebra of the configuration space manifold in the simplest case. It remains an intriguing open question what kind of theories result in, e.g., replacing the Poisson brackets by star commutators for suitable (graded) star products. 

\subsection*{Acknowledgements}
A.D. wants to thank the COST Action MP 1405 (Quantum Structure of Spacetime) for financial support. The authors want to thank Jim Stasheff and Paolo Aschieri for discussions.

\bibliography{bigone}

\bibliographystyle{jhep}

\end{document}